\def\msun{\hbox{\rm M$_{\odot}$}}

\def\HI{\hbox{\rm H\,{\sc i}}\ }

\def\HI{H{\small I}\ }
\def\simgt{\lower.5ex\hbox{$\; \buildrel > \over \sim \;$}}
\def\simlt{\lower.5ex\hbox{$\; \buildrel < \over \sim \;$}}

\documentclass[useAMS,usenatbib]{mn2e}

\usepackage{graphicx}

\title[Truncations in stellar discs]
{Radial truncations in stellar discs in galaxies}
\author[M. Kregel \&  P.~C. van der Kruit]
  {M. Kregel$^1$ \&
  P.~C. van~der~Kruit$^1$\thanks{E-mail: vdkruit@astro.rug.nl}\\
  $^1$Kapteyn Astronomical Institute, University of Groningen,
  P.O.Box 800, 9700AV Groningen, the Netherlands\\
  }
\begin{document}

\date{Accepted. Received.}

\pagerange{\pageref{firstpage}--\pageref{lastpage}} \pubyear{2004}

\label{firstpage}

\maketitle

\begin{abstract}
We discuss the possible origin of the radial truncations in stellar 
discs, using measurements that we presented 
in an earlier paper (\citealt*{KKG02}). A tentative
  correlation is found with the de-projected face-on central surface
  brightness; lower surface brightness discs tend to have a smaller
  truncation radius in units of scalelength. 
This and our earlier finding 
that in smaller spirals the truncation tends to
occur at a larger number of scalelengths are best reproduced when the
truncation is caused by a constant gas density threshhold on star formation.
\end{abstract}

\begin{keywords}
galaxies: fundamental parameters -- galaxies: spiral -- galaxies: 
structure
\end{keywords}

\section{Introduction}
\label{sec:intro}

Stellar discs have a finite size. 
In the extreme outer parts the stellar light distribution diminishes more 
steeply than the exponential decline over the main disc
and drops to low values beyond the so-called truncation radius $R_{\rm
max}$ (\citealt{K79}, \citealt*{KS81a,KS81b,KS82a}, hereafter KS1--3;
\citealt{PDLS00,SD00}; \citealt*{GKW01}, hereafter GKW; \citealt{FB01}).
This truncation or cut-off of the disc light is often directly visible
in contour maps of edge-on spirals and usually occurs at a
radius of 3 to 5 disc scalelengths \citep{K01}, although
stellar discs that extend to a much larger number of scalelengths are
certainly known to exist \citep[e.g.][]{WWGS01}. The truncation is
most easily found in edge-on spirals because of the line-of-sight
projection and the associated higher surface brightness. 

In less
inclined spirals, for which azimuthally averaged radial light profiles
are routinely studied, the non-axisymmetric component (e.g.\ spiral
structure, lopsidedness) can smooth out a truncation present in the
old disc. This effect was first noted by \citet{K88}. For 16 face-on
spirals, of which 15 did not show any sign of a truncation in the
azimuthally averaged light profile, he found that the three outermost
isophotes were much more closely spaced than the inner ones, providing
clear evidence for the truncation. The Milky Way disc probably also shows a
truncation, with recent estimates of $R_{\rm max}$ ranging from 10--15
kpc based on near-infrared star counts \citep{R96} and the near- and
far-infrared sky survey of the {\it COBE}/DIRBE instrument
\citep{F98,DS01}. Recently, wide-field surveys  have also revealed
truncations in the Local Group galaxies NGC 2403 \citep{D03} and NGC
3109 \citep{DBL03}. 

The origin of the truncation of the stellar disc is still unclear and
could possibly result from a number of physical scenarios \citep{K01,F02}.
For example, if there has been no major re-distribution of the disc
angular momentum during its formation and evolution, the truncation
may reflect the maximum specific angular momentum of the protogalaxy
\citep{K87}. This would imply that the H\,{\sc i} extending beyond the
stellar disc \citep[e.g.][]{BR97} has been accreted, and that the
truncation is associated with a small drop in the rotation curve (KS3;
\citealt{B96}). Another interesting possibility is that the truncation
corresponds to the radius at which the gas density drops below a
threshold density necessary for star formation (\citealt{FE80}; KS3;
\citealt{Ken89}; \citealt{S04}).

In \citet*[hereafter KKG]{KKG02} we fitted truncation radii to the
photometry of 34 edge-on galaxies from the sample of \citet*{G97, G98}.
The selection criteria and properties of this sample are
summarized in KKG (sect. 2 and table~1). The sample covers a
large range in rotation velocity ($v_{\rm max}$ = 50--400
km s$^{-1}$) and is dominated by spirals of intermediate- to late-type.
Of the 34 galaxies we found that at least 20 have truncated discs. The
truncation radii $R_{\rm max}$ have been listed in Table 5 in KKG. Here
we further compare the observed truncation radii to those predicted by the
scenarios proposed for the origin of the truncation, and specifically
address the behaviour with disc surface brightness (see also
\citealt{K03}).

\section{The truncation radius versus disc parameters}

\begin{figure*}
\begin{center}
\includegraphics[width=140mm]{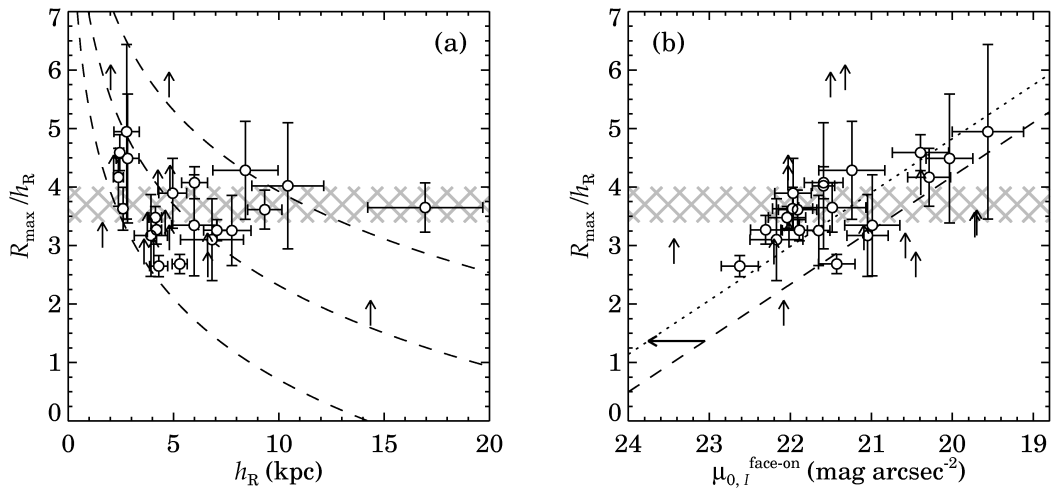}
        \caption{(a) -- $R_{\rm max}/h_{\rm R}$ versus disc
        scalelength. The arrows indicate lower limits for galaxies 
        for which no $R_{\rm max}$ could be determined. The
        cross-hatched region shows the prediction of the collapse
        model for disc galaxy formation (see text). The dashed lines
        show the predicted threshold radii for star formation
        \protect{\citep{S04}} for disc masses of, from bottom to top,
        7.5 $\times$ 10$^{9}$, 5 $\times$ 7.5 10$^{9}$ and 25 $\times$ 
        7.5 10$^{9}$ \msun. (b) -- $R_{\rm max}/h_{\rm R}$ versus face-on
        central surface brightness in the {\it I}-band. The arrows and 
         the cross-hatched region is as in
        (a). The dashed line shows the constant star-formation threshold
        prediction. The shifted dotted line is obtained when allowing
        for an underestimation of the surface brightnesses by 0.7 mag
        (see text).}
\label{fig:trunc2}
\end{center}
\end{figure*}

We show the distribution of $R_{\rm max}/h_{\rm R}$ versus scalelength
in Fig.~\ref{fig:trunc2}a. For those galaxies for which no truncation
was found, the lower limit is shown (arrows). Figure~\ref{fig:trunc2}a
appears to show a subtle increase in $R_{\rm max}/h_{\rm R}$ towards
small scalelengths; the average ratio for the spirals with $h_{\rm R}
<$ 4 kpc is 4.4, more than one standard deviation higher than that of
the entire sample. This increase of $R_{\rm max}/h_{\rm R}$ may be
related to its decrease found at very large scalelengths by
\citet{PDL00}. However, the reality of this feature is not entirely
clear considering the modest sample size and the selection effect
against galaxies of small physical sizes and low surface
brightness. We are, for example, still missing small low surface
brightness galaxies, which may have entirely different $R_{\rm
max}/h_{\rm R}$. While our view of the distribution of $R_{\rm
max}/h_{\rm R}$ is certainly obscured by this selection effect,
additional information can be obtained from face-on samples. We have
estimated a lower limit to $R_{\rm max}$ for the spirals in the
face-on sample of \citet{JK94}, a sample which is dominated by spirals
of small scalelength. By taking the lowest contour in their {\it
R}-band contour plots as a lower limit to $R_{\rm max}$, we find an
average $\left(R_{\rm max}/h_{\rm R}\right)_{\rm min} =$ 4.0$\pm$1.1
(1$\sigma$). This lower limit and the tentative increase of $R_{\rm
max}/h_{\rm R}$ towards small scalelength spirals suggest that
the ratio of
truncation radius to disc scalelength in small scalelength 
nearby spirals is at least four.

Figure~\ref{fig:trunc2}b shows the dimensionless truncation radius
versus the de-projected face-on central surface brightness (corrected
for Galactic extinction as in \citealt{G98}). The ratio $R_{\rm
max}/h_{\rm R}$ appears to be correlated with surface brightness in
the sense that lower surface brightness discs tend to truncate at a
smaller number of scalelengths. A Spearman rank correlation test
appears conclusive, yielding a correlation coefficient of 0.61 (or a
confidence level greater than 99\%). However, the errors are 
substantial in many cases and not entirely random, so that the test 
could have given a false positive result. Note
that the face-on surface brightnesses are probably underestimated, by
$\sim$ 0.7 mag according to a comparison between the Tully-Fisher
relations of edge-on and face-on spirals (\citealt*{KKF04}; hereafter
KKF).
The large lower limits for ESO 340-G08 and ESO 555-G36 are puzzling.
The discs of these spirals appear to be far
more extended than the norm.

Figures~\ref{fig:trunc2}a \& b combined suggest that in small
scalelength, high surface brightness spirals the truncation occurs at
at least four scalelengths. This is important, because small spirals
($h_{\rm R} \simlt$ 4 kpc) are the most numerous in the local
Universe (\citealt{K87}; \citealt{JL00}, KKG).
As another corollary, the range in face-on surface brightness among
spirals is narrower at $R_{\rm max}$ than at the disc centre. If the
discs in the current sample are exponential out to $R_{\rm max}$
then the distribution of {\it I}-band face-on surface brightnesses at
$R_{\rm max}$ has an average of 25.3 mag arcsec$^{-2}$ and a 1$\sigma$
scatter of 0.6 mag arcsec$^{-2}$. This dispersion is considerably
smaller than the 1.1 mag arcsec$^{-2}$ dispersion in the central
surface brightness. This also implies that in the face-on view,
the truncation is just as easily (or laboriously) detected in LSB as
in HSB spirals.

\section{Comparison with hypotheses for the origin of the truncations}

We can compare the observed trends to the predictions of the
analytical collapse model of disc galaxy formation
(\citealt*{FE80,G82,DSS97}, hereafter DSS97).  If protogalaxies are relatively
sharp-edged, then the collapse theory also predicts the outermost
radius of the baryonic disc \citep{K87}. This radius corresponds to
the material with the highest specific angular momentum in the
protogalaxy. To quantify this prediction we calculated model surface
density profiles using the method of DSS97 for a range in spin
parameter and total mass of the protogalaxy. We assume a constant
baryonic mass fraction $F =$ 0.10 and a constant efficiency for
turning the baryons into stars over the age of the disc at $(M/L)_{\rm
disc} =$ 2 (see KKF). 
The scalelengths were derived using a method similar to
the `marking-the-disc' method \citep{F70}, and the outermost radii
were obtained by taking the radius at which the density drops to
zero. The result is indicated by the cross-hatched region in
Figs.~\ref{fig:trunc2}.

In the collapse theory, both the outermost radius and the scalelength
of the baryonic proto-disc increase with the mass and angular momentum
of the protogalaxy such that their ratio remains approximately
constant at 3--4. Taking a different or non-constant baryon mass
fraction does not significantly change this result. The predicted
$R_{\rm max}/h_{\rm R}$ is slightly smaller than the ratio of 4.5
predicted by \citet{K87} based on a comparison of the angular momentum
distribution of an exponential disc with that of a uniformly rotating,
uniform sphere with $\lambda =$ 0.07. Although the prediction of the
collapse model roughly coincides with the average observed $R_{\rm
max}/h_{\rm R}$, it can not explain the existence of discs which
extend to a relatively large or a small number of scalelengths. In
particular, it does not predict an increase of $R_{\rm max}/h_{\rm R}$
toward small scalelengths (Fig.~\ref{fig:trunc2}a), or a decrease
towards low surface brightness discs (Fig.~\ref{fig:trunc2}b). Taking
a different halo density profile and/or angular momentum profile could
change the value of $R_{\rm max}/h_{\rm R}$, but not its constancy.
Including additional prescriptions for mass accretion, star formation
and supernova feedback can lead to $R_{\rm max}/h_{\rm R}$ ratios that
do change as a function of central surface brightness, but with the
opposite sign \citep{vdB01}. Perhaps taking a range in angular
momentum profiles, as suggested by N-body simulations in the
$\Lambda$CDM cosmology \citep{BDK01}, or re-distributing the angular
momentum during and/or after the collapse could solve the
discrepancy.
 
Alternatively, the stellar disc truncation may be caused by the
inhibition of widespread star formation below a critical gas surface
density (\citealt{FE80}; KS3; \citealt{Ken89}), 
{\it if} the corresponding critical radius
is approximately constant over time. This star formation threshold is
suspected to be related to the stability of the gas disc. 
Recently, \citet{S04} made a prediction for
the threshold radius based on simulations of the thermal and
ionization structure of the gaseous discs assembled in the galaxy
formation model of \citet{MMW98}. In these simulations the transition
to the cold ISM phase is responsible for the onset of local
gravitational instability that triggers star formation. This
transition to the cold phase is independent of the shape of the
rotation curve and occurs at a critical gas surface density, which for
reasonable values of the gas fraction, turbulence, metallicity and
the UV radiation intensity attains values in the range $\Sigma_{\rm c}
\sim $ 3--10 \msun\, pc$^{-2}$ \citep{S04}. For a disc with an exponential
surface density profile, he finds:

\begin{equation}
\frac{R_{\rm max}}{h_{\rm R}} = \ln \frac{M_{\rm disc}}{2\pi h_{\rm R}^{2}\Sigma_{\rm c}},
\label{eqn:trunc}
\end{equation}

\noindent
where $M_{\rm disc}$ is the disc mass (gas and stars) and $h_{\rm R}$
is the mass scalelength. It is easy to show that when the gas mass is
negligible this prediction reduces to:

\begin{equation}
\frac{R_{\rm max}}{h_{\rm R}} = \ln \frac{(M/L)_{\star}
\mu_{0}}{\Sigma_{\rm c}}
\label{eqn:trunc_nogas}
\end{equation}

\noindent
with $(M/L)_{\star}$ the stellar mass-to-light ratio and $\mu_{0}$ its
central surface brightness (linear units).

\begin{figure}
\begin{center}
\includegraphics[width=70mm]{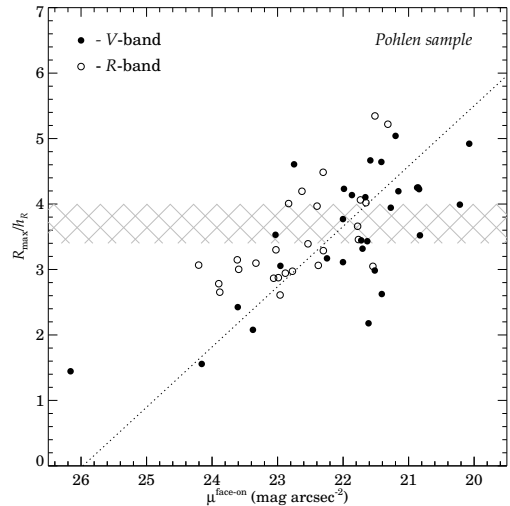}
        \caption{$R_{\rm max}/h_{\rm R}$ versus face-on central
        surface brightness for the {\it V} and {\it R}-band results of
        Pohlen for the sharply truncated disc model (\citealt{P01}). 
        The face-on surface brightnesses were calculated
        assuming a exponential vertical luminosity distribution. As in
        Fig.~1b, the cross-hatched region shows the prediction of the
        collapse model and the dotted line shows the prediction of the
        star-formation treshold model, with the except that here the
        $(M/L)_{\star}$ for the threshold model is arbitrary.}
\label{fig:pohlen}
\end{center}
\end{figure}

The predicted threshold radii according to Eqn.~\ref{eqn:trunc} are
shown in Fig.~\ref{fig:trunc2}a for three disc masses and Schaye's
(\citeyear{S04}) fiducial critical surface density $\Sigma_{\rm c} =$
5.9 \msun\, pc$^{-2}$ ($N_{\rm H}$ = 5.6 10$^{20}$ cm$^{-2}$). For the
adopted disc masses, the threshold model brackets the observations.
Interestingly, the model predicts an increase in $R_{\rm max}/h_{\rm
R}$ towards small scalelengths. Discs with a higher central surface
density form stars out to a larger radius in terms of scalelengths
before reaching the critical density. Since, for constant total disc
mass, a higher surface density disc has a smaller scalelength, it
follows that smaller scalelength discs have larger $R_{\rm max}/h_{\rm
R}$. This anti-correlation, which is similar to the observed trend in
the \citet{PDL00} sample \citep{S04}, is also in accordance with the
present observations.

Motivated by the observation that the \HI gas fraction in the galaxies
is small (\citealt*{KKB04}, hereafter KKB), we show 
Eqn.~\ref{eqn:trunc_nogas} in
Fig.~\ref{fig:trunc2}b assuming $\Sigma_{\rm  c} =$ 5.9 \msun
pc$^{-2}$ and $(M/L)_{\star} =$ 2 (constant among galaxies). Lower
surface brightness discs are predicted to be less extended, with a
slope in agreement with the observed trend. The match is better
if we take into account that the inferred disc central surface
brightnesses are fainter by about 0.7 magnitudes compared to their
face-on counterparts (KKF). The scatter and outliers
may be explained as being due to a non-constancy in $\Sigma_{\rm c}$,
e.g.\ due to a varying UV radiation intensity or metallicity, or
due to a variation of $(M/L)_{\star}$. Altogether, the observations
are consistent with a `critical surface brightness' in the range
$\Sigma_{\rm c}/(M/L)_{\star} =$ 1.5--4 L$_{\sun,\it I}$ pc$^{-2}$.
Saying it differently, for $\Sigma_{\rm c} =$ 5.9 \msun pc$^{-2}$ the
stellar mass-to-light ratio is $(M/L)_{\star} =$ 4--1.5
M$_{\sun}$/L$_{\sun,\it I}$. The lower part of this range is
reasonable, both from the perspective of stellar population synthesis
\citep{BJ00} and observations of the stellar velocity dispersions in
galaxy discs \citep{B97}.

Both models can also be compared to the truncation analysis performed
by \citet{P01} in the {\it V} and the {\it R} bands (Fig.~\ref{fig:pohlen}).
Although obtained using a different method, these data show a trend
very similar to our {\it I}-band result (Fig.~\ref{fig:trunc2}) and
to the star-formation treshold model (dotted line). Lower surface
brightness discs tend to be less extended in terms of $R_{\rm max}$.

It has long been known that viscosity driven angular momentum
re-distribution within a star-forming gas disc may drive the
resulting stellar disc towards an exponential profile
\citep{LP87}. Depending on the details of star formation, this
process may yield a stellar disc with a well-defined edge that
advances radially outward with time \citep{FC01}. Unfortunately, the
theory has not yet been explicitly investigated with respect to this
truncation radius. The theory of stochastic self-propagating star
formation predicts $R_{\rm max}/h_{\rm R}$ = 4 \citep{SSE84} for
flat rotation curves. This prediction is similar to that of the
collapse model, leading to the same difficulties
of predicting no dependence on surface brightness.

\section{Conclusions}
\label{sec:conclusions}

At least 20 galaxies in our sample of 34
have truncated stellar discs, displaying a tight
relation between disc scalelength and truncation radius.
The stellar disc
edge seems to occur at a larger number of scalelengths in galaxies
with a smaller scalelength, in agreement with the decrease of $R_{\rm
max}/h_{\rm R}$ towards large scalelengths reported by \citet{PDL00}.
In addition, we observe a tentative correlation between the ratio
$R_{\rm max}/h_{\rm R}$ and the de-projected face-on central surface
brightness of the disc. These observations appear to have two
important implications. First, high surface brightness spirals with
small scalelengths, which are the most numerous spirals in the local
Universe \citep{JL00}, have an $R_{\rm max}/h_{\rm R}$ of at least
four. Secondly, the face-on disc surface brightness at the truncation
radius is roughly constant among galaxies.

The observed truncation radii were compared to the predictions of
several theories proposed for its origin. The data are best reproduced
by the constant star-formation threshold model for the truncation
(\citealt{S04}). In particular,
this model is able to match the correlation between $R_{\rm
max}/h_{\rm R}$ and surface brightness for reasonable stellar
mass-to-light ratios. The collapse model for disc galaxy formation
(\citealt{K87}; DSS97), in which the truncation corresponds to the
maximum angular momentum in the proto-galaxy before disc collapse with
detailed conservation of angular momentum,
provides a significantly poorer match to the observations.
This model would require some amount of re-distribution of
angular momentum during and/or after the collapse.

\section*{Acknowledgments}

We thank the referee for comments and suggestions.

\bsp

\label{lastpage}

\end{document}